\pdfoutput=1
\documentclass[review]{elsarticle}

\usepackage[colorlinks=true, pdfstartview=FitV, linkcolor=blue, citecolor=blue, urlcolor=blue]{hyperref}
\usepackage{amsmath,amsthm}
\usepackage{color}
\usepackage{ulem}

\newcommand{\tr}{\mathrm{tr}}
\newcommand{\N}{\mathcal{N}}
\newcommand{\A}{A}
\newcommand{\D}{\mathcal{D}}
\renewcommand{\H}{\mathcal{H}}
\renewcommand{\P}{\mathcal{P}}

\begin{document}

\begin{frontmatter}

\title{Abelian and non-Abelian Berry curvatures \\ in lattice QCD}

\author{Shi Pu}
\author{Arata Yamamoto\corref{co}}
\cortext[co]{Corresponding author}
\ead{arayamamoto@nt.phys.s.u-tokyo.ac.jp}

\address{Department of Physics, The University of Tokyo, Tokyo 113-0033, Japan}

\begin{abstract}
We studied the Berry curvature of the massive Dirac fermion in 3+1 dimensions.
For the non-interacting Dirac fermion, the Berry curvature is non-Abelian because of the degeneracy of positive and negative helicity modes.
We calculated the non-Abelian Berry curvature analytically and numerically.
For the interacting Dirac fermion in QCD, the degeneracy is lost because gluons carry helicity and color charge.
We calculated the Abelian Berry curvature in lattice QCD.
\end{abstract}


\end{frontmatter}

\section{Introduction}

The geometric phase known as the Berry phase \cite{Berry1984} is universal in physics.
The Berry phase is defined by the change of wave functions in a cyclic adiabatic evolution.
During an adiabatic evolution, at each time, a system approximately stays at the
eigenstate $|n(t)\rangle$ of the temporal Hamiltonian $H(t)$. The variation of those temporal eigenstates will contribute to
wave functions as an extra phase factor $\gamma = \int_\mathcal{C} dt \langle n(t)|i\partial_t|n(t) \rangle$.
In a cyclic adiabatic case, this phase factor cannot be removed by choosing different paths $\mathcal{C}$ in parameter space, and, therefore is gauge-invariant.
The Berry phase is an analogy to the Aharonov-Bohm phase of ordinary electromagnetic fields.
One can also introduce the Berry connection and the Berry curvature corresponding to a vector potential and magnetic field, respectively.
A well-known manifestation of the Berry phase is the modification of the equations of motion of charged particles under electromagnetic fields \cite{Xiao:2005eif,Duval2006}.
(See also the reviews \cite{Xiao2010a,Chang2008} and references therein.)  

One novel consequence of the coupling between the Berry curvature and a magnetic field is 
a quantum non-dissipative transport effect,
named chiral magnetic effect (CME) \cite{Kharzeev:2007jp,Fukushima2008a,Kharzeev:2009pj}. 
A charge current can be induced by magnetic fields in the presence of imbalanced chiralities of Weyl fermions, which was confirmed in quantum field theory \cite{Fukushima2008a,Kharzeev:2009pj}, in relativistic hydrodynamics \cite{Son:2009tf,Pu:2010as,Sadofyev:2010pr,Kharzeev:2011ds}, in the holographic models \cite{Erdmenger2009,Torabian2009a,Banerjee2011,Landsteiner2011}, and in lattice quantum chromodynamics (QCD) simulations \cite{Buividovich:2009zzb,Buividovich:2009wi,Buividovich2010,Yamamoto:2011gk,Yamamoto:2011ks,Bali:2014vja}.
Those studies have drawn lots of attentions both in high energy physics and in condensed matter physics.
In high energy physics, the CME is applied to understand the azimuthal charged-particle correlations in heavy ion collisions \cite{Abelev2009}.
In condensed matter physics, the CME may lead to the negative magnetoresistance of Weyl metals and modify
electric conductivity \cite{Son2013a,Li2016b}. 
More references and relevant studies can be found in the reviews
\cite{Basar:2012gm,Fukushima:2012vr,Kharzeev:2012ph,Kharzeev:2015znc}.

More deep connection between the CME and the Berry curvature shows up when people build up the kinetic theory of Weyl fermions with quantum corrections, which is called chiral kinetic theory; see Refs.~\cite{Son:2012wh,Son:2012zy} for the Hamiltonian formalism, Refs.~\cite{Stephanov:2012ki,Chen:2014cla,Chen:2013iga} for the path integrals, Refs.~\cite{Chen:2012ca,Gao:2017rgi} 
for the Wigner functions near the equilibrium and Refs.~\cite{Hidaka:2016yjf,Ebihara:2017suq} out-off equilibrium, and also Refs.~\cite{Mueller:2017lzw,Mueller:2017arw} for the world-line formalism.
In the chiral kinetic theory, the Berry curvature involving the spin effects of Weyl fermions couples to external magnetic fields.
The pole of the Berry curvature at zero momentum plays the role of a magnetic monopole.
This finally contributes to a charge current and the non-conservation of axial charges, which are exactly the CME and chiral anomaly.  
Recently, people have also extended the discussion to some non-linear responses of electromagnetic fields \cite{Chen:2016xtg,Gorbar:2016qfh,Gorbar:2017cwv,Gorbar:2017toh,Hidaka:2017auj}.

For massive Dirac fermions, the Hamiltonian is not diagonal and two chirality states are not independent due to a mass term.
The lowest energy level obtained by diagonalizing the Hamiltonian has the two-fold degeneracy corresponding to two spins or helicities.
The particle has SU(2) symmetry, i.e. it can freely switch from one of the lowest energy levels to the other.
Therefore the Berry curvature become non-Abelian like non-Abelian gauge theory \cite{Zee1984}.
Such a non-Abelian Berry curvature will modify dynamic evolution of Dirac wave-packets \cite{Chang2008,Chuu2010533} and then has been used to build up the quantum 
kinetic theory for massive fermions \cite{Chen:2013iga,Stone2013}\footnote{Note that the authors in 
Ref.~\cite{Stone:2014fja} have introduced a covariant Berry connection in the massive case.
However, its definition is not the same as what we discuss in this work.}.

From the viewpoint of quantum field theory, a natural question comes out:
what happens for the massive Dirac fermions in QCD when the gluon interaction is involving?
Although the non-interacting limit of the Berry curvature is well understood, quantum corrections and interaction effects need more studies.
One promising way to deal with the gluon interaction is lattice QCD. 
In this work, our main propose is to calculate the Berry curvature for massive fermions with the gluon interaction.
We will follow the lattice calculation of the Abelian Berry curvature formulated in the previous work \cite{Yamamoto:2016rfr} and generalize it to the non-Abelian cases.
One possible application of this work is the future simulation of chiral kinetic theory in relativistic heavy ion collisions.
Back to 90's, many pioneer works have revealed the relation between geometric phases and the chiral anomaly in the context of chiral fermions on the lattice \cite{NARAYANAN199362,Narayanan:1994gw,RANDJBARDAEMI1995543,RandjbarDaemi:1995cq,RandjbarDaemi:1995gy,RandjbarDaemi:1995mj,RANDJBARDAEMI1997134,PhysRevD.59.085006}, although these works are not directly related to our work. 
For recent related topics, also see several first-principle calculations in condensed matter physics \cite{Gradhand2012,Yamamoto:2016zpx}.

In Sec.~\ref{secB}, we introduce the basic equations and the lattice calculation of the Abelian and non-Abelian Berry curvatures.
In Sec.~\ref{secF}, we calculate the non-Abelian Berry curvature of non-interacting massive Dirac fermions.
In Sec.~\ref{secQCD}, we calculate the Abelian Berry curvature in quenched lattice QCD.

\section{Berry curvature}
\label{secB}

Suppose that the ground state of a fermion is $N$-fold degenerate,
\begin{equation}
 \Phi_n (p) \quad (n = 1,2, \cdots, N)
,
\end{equation}
where $p$ is the spatial momentum of the fermion.
All the ground states satisfy the orthonormal condition $\Phi^\dagger_m \Phi_n = \delta_{mn}$.
The Berry connection is given by 
\begin{equation}
 \A_j = \sum_{\alpha=0}^{N^2-1} T^\alpha \A_j^\alpha
,
\end{equation}
where $T^\alpha$ is the Lie algebra of U(N).
The matrix elements are defined by
\begin{equation}
\label{eqA}
[\A_j (p)]_{mn} = -i \Phi^\dagger_m(p) \frac{\partial}{\partial p_j} \Phi_n(p)
.
\end{equation}
The Berry connection is Abelian for a non-degenerate ground state ($N=1$) and non-Abelian for degenerate ground states ($N\ge 2$).
Note that the index $j$ runs only spatial dimensions, i.e., $j=1,2,3$ in 3+1 dimensions.
The Berry curvature is defined by
\begin{equation}
\label{eqOmega}
 \Omega_{ij}(p) = \frac{\partial}{\partial p_i} \A_j (p) - \frac{\partial}{\partial p_j} \A_i (p) - i[\A_i (p),\A_j (p)]
.
\end{equation}
The Berry curvature is also the generators of U(N).
At the leading order of $\Omega_{ij}$, the field strength
\begin{equation}
 \sum_{i,j} \frac{1}{2} \tr[ \Omega_{ij}(p) \Omega_{ij}(p) ] = \sum_{\alpha} \sum_{i,j} \frac{1}{4} \Omega_{ij}^\alpha(p) \Omega_{ij}^\alpha(p)
\end{equation}
is the unique term under Lorentz symmetry and gauge symmetry.
At higher orders, there are many possible terms.
For example, at the next-to-leading order, there are $\sum_{i,j,k} \tr[ (\D_i \Omega_{jk})(\D_i \Omega_{jk}) ]$, $\sum_{i,j,k} \tr[ (\D_i \Omega_{ik})(\D_j \Omega_{jk}) ]$, and $\sum_{i,j,k} \tr[ \Omega_{ij} \Omega_{jk} \Omega_{ki} ]$, where $\D_i$ is the covariant derivative in the U(N) adjoint representation.
These terms can be constructed in lattice gauge theory \cite{Kanazawa:2014fla}.

The lattice QCD calculation of the Abelian Berry curvature has been formulated \cite{Yamamoto:2016rfr}.
We here extend it to the non-Abelian one.
The lattice calculation is performed in a finite volume.
Momentum space is also discretized in finite coordinate space.
The coordinate-space lattice spacing $a$ and the momentum-space lattice spacing $\tilde{a}$ are related by $\tilde{a} = 2\pi/N_sa$, where $N_s$ is the lattice size.
The Berry connection is described by the gauge field on the momentum-space lattice \cite{2005JPSJ...74.1674F}.
The Abelian Berry curvature is described by Abelian lattice gauge theory, like lattice QED, and the non-Abelian Berry curvature is described by non-Abelian lattice gauge theory, like lattice QCD.

In the Abelian case ($N=1$), the ground state is uniquely obtained by the large Euclidean-time limit $\tau \to \infty$ as
\begin{equation}
\label{eqGSU1}
 \Phi(p) = \frac{1}{\N(p)} \lim_{\tau \to \infty} D^{-1}(p,\tau) \phi_{\rm init} (p)
.
\end{equation}
The factor $\N$ is the normalization constant and $D^{-1}$ is the fermion propagator in momentum space.
The initial state $\phi_{\rm init}$ is arbitrary as long as it has nonzero overlap with the ground state.
The Berry link variable is given by
\begin{equation}
\label{eqUU1}
 U_j(p) = \frac{ \Phi^\dagger(p) \Phi(p + \tilde{j}) }{|\Phi^\dagger(p) \Phi(p + \tilde{j}) |} = \exp \left[ i \tilde{a} \A_j(p) \right]
.
\end{equation}
The symbol $\tilde{j}$ is the unit lattice vector in $p_j$ direction.
In the non-Abelian cases ($N\ge 2$), the ground state is not unique.
We introduce the projection operators $\P_n$ $(n=1,2,\cdots, N)$ to obtain each of the degenerate ground states.
The projection operators commute with the Hamiltonian, $[\P_n, \H]=0$, otherwise, projected states are not ground states.
The ground states are obtained by
\begin{equation}
\label{eqGS}
 \Phi_n(p) = \frac{1}{\N_n(p)} \lim_{\tau \to \infty} \P_n(p) D^{-1}(p,\tau) \phi_{\rm init} (p)
.
\end{equation}
Naively, one would take the Berry link variable
\begin{equation}
\label{eqV}
 [V_j(p)]_{mn} = \frac{ \Phi^\dagger_m(p) \Phi_n(p + \tilde{j}) }{|\det \Phi^\dagger(p) \Phi(p + \tilde{j}) |^{1/N}}
\end{equation}
by the analogy to the Abelian case \eqref{eqUU1}.
The matrix $V_j(p)$ is, however, not U(N) at $\tilde{a} \neq 0$.
It becomes U(N) only in the limit of $\tilde{a} \to 0$.
We construct the U(N) Berry link variable 
\begin{equation}
\label{eqU}
 U_j(p) = \exp \left[ i \tilde{a} \A_j(p) \right]
\end{equation}
by maximizing the overlap
\begin{equation}
{\rm Re}\, \tr \left[ V_j^\dagger(p) U_j(p) \right] 
.
\end{equation}
We define the Abelian or non-Abelian Berry plaquette
\begin{equation}
\label{eqPlaq}
 U_{ij}(p) 
= U_i(p) U_j(p+\tilde{i}) U^\dagger_i(p+\tilde{j}) U^\dagger_j(p)
= \exp \left[ i\tilde{a}^2 \Omega_{ij}(p) \right]
\end{equation}
from the Berry link variable \eqref{eqUU1} or \eqref{eqU}.
The plaquette is the fundamental quantity to calculate the field strength in lattice gauge theory.

\section{Free Dirac fermion}
\label{secF}

We first study the non-interacting Dirac fermion.
In 3+1 dimensions, the Dirac fermion is a four-component spinor.
There are four solutions: particle states with positive and negative helicity and anti-particle states with positive and negative helicity.
The positive and negative helicity states are degenerate.
Thus, we have the U(2) non-Abelian Berry curvature.
This is a specialty in 3+1 dimensions different from 1+1 or 2+1 dimensions.
(When the Dirac fermion is a two-component spinor, there are only two solutions: a positive-energy state and a negative-energy state.
We have the U(1) Abelian Berry curvature.)

In a continuum limit and in an infinite volume, we can analytically calculate the Abelian and non-Abelian Berry curvatures. 
The field strengths of the particle states are
\begin{equation}
\label{eqFanalytic}
\begin{split}
& \sum_{i,j} \frac{1}{4} \Omega_{ij}^0(p) \Omega_{ij}^0(p) = 0
\\
& \sum_{a\ne 0} \sum_{i,j} \frac{1}{4} \Omega_{ij}^a(p) \Omega_{ij}^a(p) = \frac{E^2 + 2m^2}{4E^6}
\end{split}
\end{equation}
with $E= \sqrt{p^2 + m^2}$.
The Abelian component is zero because of the cancellation between positive and negative helicity states.
The non-Abelian component is nonzero and finite.
More details can be found in Appendix.
Chemical potentials are assumed to be zero.
If a quark chemical potential $\mu$ is nonzero, the energy shifts as $E \to E-\mu$.
This breaks particle-anti-particle symmetry but keeps helicity symmetry.
If a chiral chemical potential $\mu_5$ is nonzero, the eigenstate of helicity operator shifts as $\pm p \to \pm p - \mu_5$.
This breaks helicity symmetry, and thus the Berry curvature turns from U(2) to U(1).
The Abelian field strength becomes nonzero because the cancellation does not occur.

We performed the lattice calculation of the massive Dirac fermion to demonstrate the non-Abelian formulation in Sec.~\ref{secB}.
We used the helicity projection operator
\begin{equation}
\label{eqProj}
\P_{\pm}=\frac{1}{2}\left\{ 1\pm\frac{1}{\sqrt{\sum_i \sin^2(p_ia)}}
\left(
\begin{array}{cc}
\sum_{j}\sigma_{j} \sin(p_{j}a) & 0\\
0 & \sum_{j}\sigma_{j} \sin(p_{j}a)
\end{array}
\right)\right\} 
\end{equation}
in the ground state calculation \eqref{eqGS}.
The lattice size is $N_s^3 \times N_\tau = 16^3 \times 32$.
We used the unimproved Wilson fermion with the bare mass $ma = 0.22$.
The Berry plaquette \eqref{eqPlaq} is replaced by the clover improved one to preserve the reflection symmetry \cite{Yamamoto:2013zwa,Yamamoto:2014vda}.

\begin{figure}[h]
\begin{minipage}{0.49\textwidth}
 \includegraphics[width=1\textwidth]{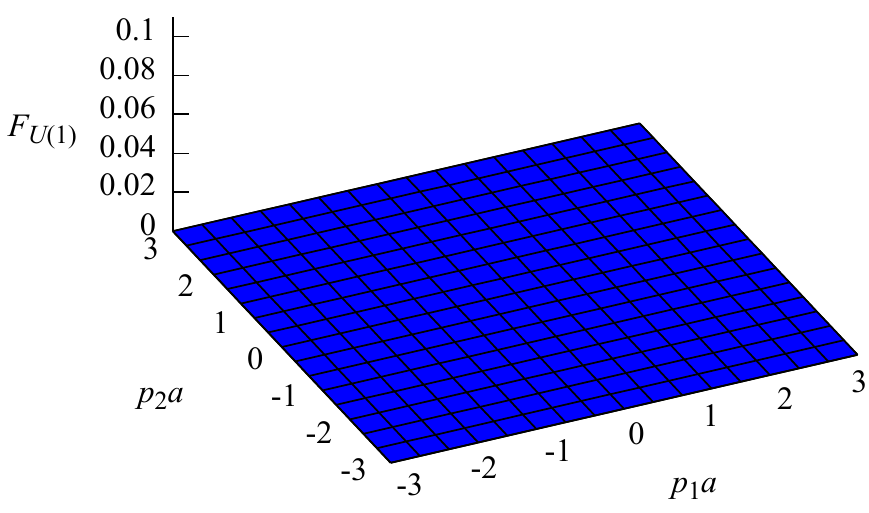}
\end{minipage}
\begin{minipage}{0.49\textwidth}
 \includegraphics[width=1\textwidth]{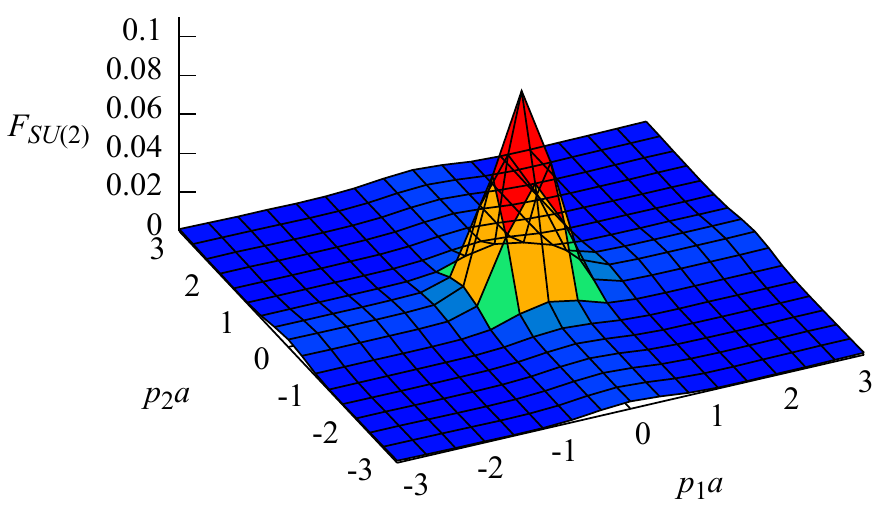}
\end{minipage}
\caption{
\label{figB0}
Berry field strength of the non-interacting Dirac fermion: $F_{U(1)}$ (left) and $F_{SU(2)}$ (right).
The data at $p_3=\tilde{a}$ are shown.
}
\end{figure}

\begin{figure}[h]
 \includegraphics[width=.8\textwidth]{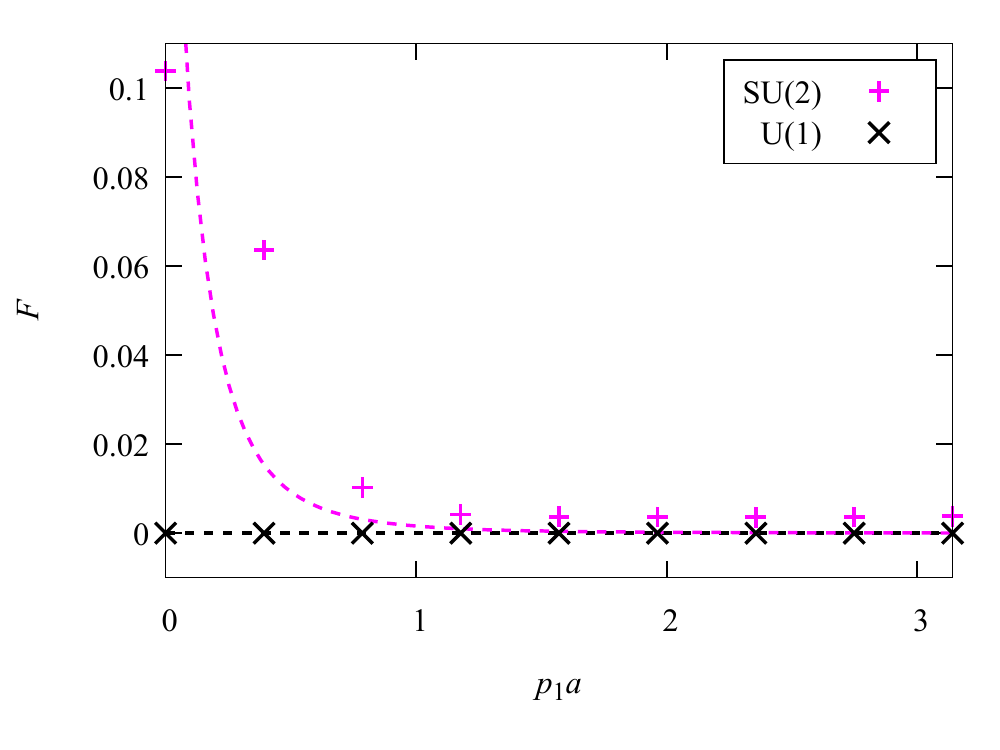}
\caption{
\label{figF0}
Berry field strength $F_{U(1)}$ and $F_{SU(2)}$ of the non-interacting Dirac fermion.
The data at $p_2=0$ and $p_3=\tilde{a}$ are shown.
The broken line is the analytical result \eqref{eqFanalytic} multiplied by $\tilde{a}^4$.
}
\end{figure}

The U(2) plaquette \eqref{eqPlaq} can be decomposed into the U(1) and SU(2) parts
\begin{equation}
\begin{split}
 r_{ij}(p) &= \{ \det U_{ij}(p) \}^{1/2} = \exp\left[i\tilde{a}^2 T^0 \Omega_{ij}^0(p) \right]
\\
 R_{ij}(p) &= \frac{U_{ij}(p)}{\{ \det U_{ij}(p) \}^{1/2}} = \exp\left[i\tilde{a}^2 \sum_{a\ne 0} T^a \Omega_{ij}^a(p) \right]
.
\end{split}
\end{equation}
The field strengths are given by
\begin{equation}
\label{eqF}
\begin{split}
 F_{U(1)} &\equiv \sum_{i,j} \{\arg r_{ij}(p)\}^2 \simeq \sum_{i,j} \frac{\tilde{a}^4}{4} \Omega_{ij}^0(p) \Omega_{ij}^0(p)
\\
 F_{SU(2)} &\equiv \sum_{i,j} \{ 2 - \tr R_{ij}(p) \} \simeq \sum_{a\ne 0} \sum_{i,j} \frac{\tilde{a}^4}{4} \Omega_{ij}^a(p) \Omega_{ij}^a(p)
.
\end{split}
\end{equation}
The approximate equality means the leading order of $\tilde{a}$.
The results are shown in Figs.~\ref{figB0} and \ref{figF0}.
Since the helicity projection operator \eqref{eqProj} is singular at $p_1=p_2=p_3=0$, the data at $p_3 = \tilde{a}$ are shown.
As seen in Fig.~\ref{figF0}, the lattice calculation in a finite volume is almost consistent with the analytical calculation in the infinite volume.
This ensures the validity of the present formulation.
The small-momentum region is modified by finite volume effect.
If the continuum momentum limit $\tilde{a} \to 0$, i.e., the infinite volume limit $N_s \to \infty$ is taken, the lattice calculation would reproduce the analytical calculation more accurately.

\section{QCD}
\label{secQCD}

Next we study the interacting Dirac fermion in QCD.
The interacting Dirac fermion has internal color space as well as spinor space.
Although the helicity and color charge of a total system are conserved, the helicity and color charge of fermions are not conserved because gluons carry them.
We cannot construct any projection operator commutable with the interacting Dirac Hamiltonian.
The non-interacting ground states mix and result in one non-degenerate ground state.
The Berry curvature is U(1)\footnote{Note that, in Ref.\cite{Chen:2013iga}, the authors have assumed the
interaction is very week. Therefore, in that case, the lowest energy level is approximately degenerate and there are non-Abelian Berry curvature. }.

We calculated the Berry curvature in quenched lattice QCD.
We adopted the lattice coupling constant $\beta = 6/g^2 = 5.9$ and the hopping parameter $\kappa = 0.1583$.
They corresponds to the lattice spacing $a \sim 0.10$ fm and a half of the $\rho$ meson mass $m_\rho a/2 \sim 0.22$ \cite{Aoki:2002fd}.
Since the fermion propagator in momentum space is gauge variant in Eq.~\eqref{eqGSU1}, we performed the SU(3) Coulomb gauge fixing of gluon fields\footnote{Do not be confused with this gauge fixing. This is the gauge fixing for the SU(3) gluon field in coordinate space, not for the U(1) Berry connection in momentum space.}.
Other conditions were the same as the non-interacting lattice calculation in Sec.~\ref{secF}.

\begin{figure}[h]
\begin{minipage}{0.49\textwidth}
 \includegraphics[width=1\textwidth]{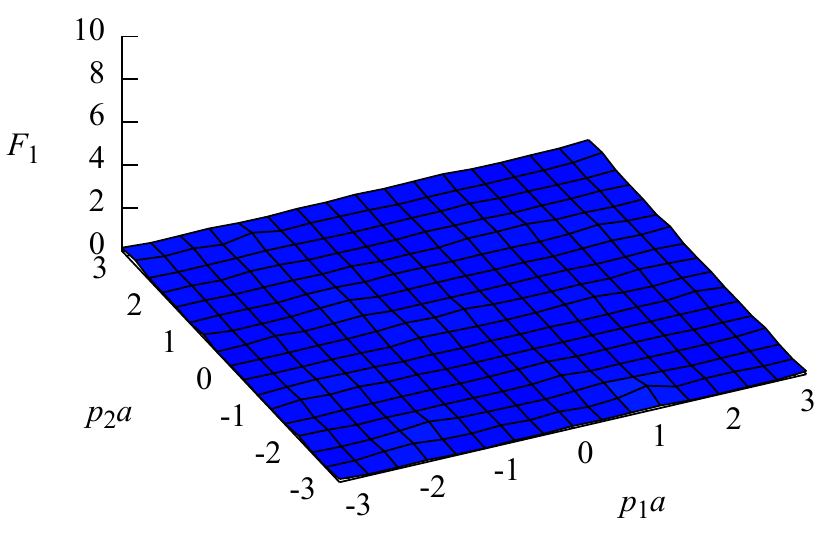}
\end{minipage}
\begin{minipage}{0.49\textwidth}
 \includegraphics[width=1\textwidth]{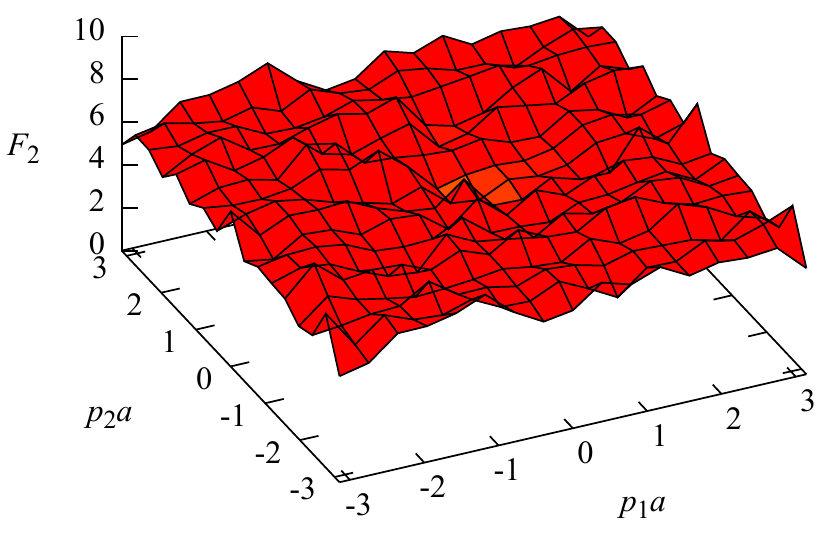}
\end{minipage}
\caption{
\label{figB1}
U(1) Berry field strength of the interacting Dirac fermion: $F_1$ (left) and $F_2$ (right).
The data at $p_3=\tilde{a}$ are shown.
}
\end{figure}

\begin{figure}[h]
 \includegraphics[width=.8\textwidth]{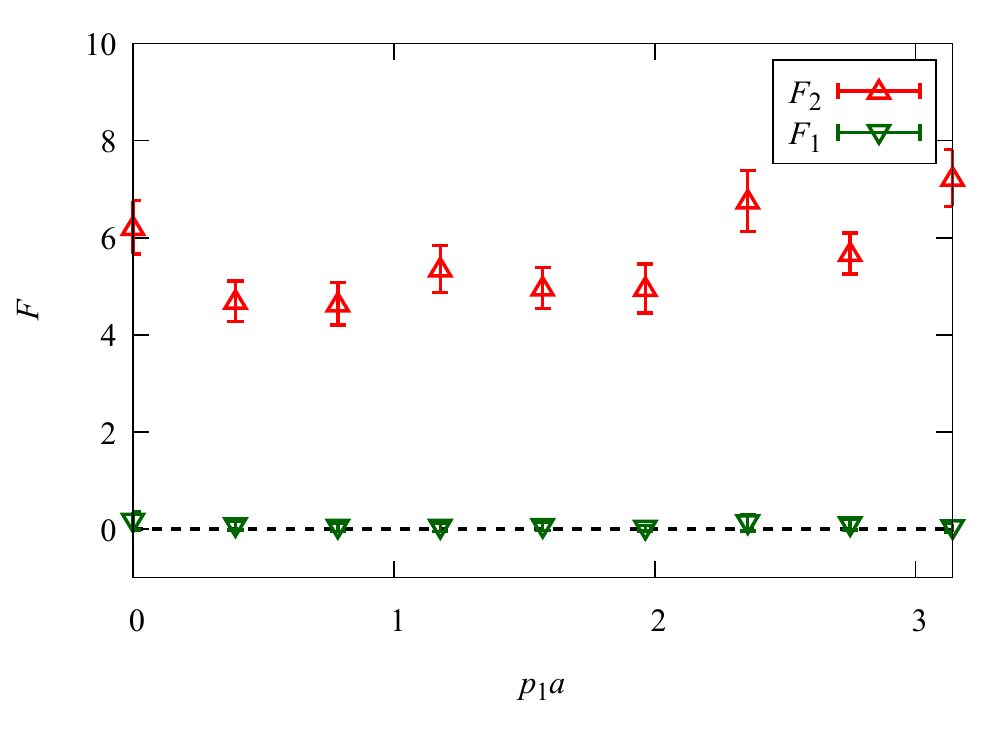}
\caption{
\label{figF1}
U(1) Berry field strength $F_1$ and $F_2$ of the interacting Dirac fermion.
The data at $p_2=p_3=0$ are shown.
}
\end{figure}

We calculated two kinds of the field strength
\begin{equation}
\begin{split}
 F_1 &\equiv \sum_{i,j} \langle \arg U_{ij}(p) \rangle^2 \simeq \sum_{i,j} \left\langle \frac{\tilde{a}^2}{2} \Omega_{ij}(p) \right\rangle^2
\end{split}
\end{equation}
and
\begin{equation}
\begin{split}
 F_2 &\equiv \sum_{i,j} \left\langle \{\arg U_{ij}(p)\}^2 \right\rangle \simeq \sum_{i,j} \left\langle \frac{\tilde{a}^4}{4} \Omega_{ij}(p) \Omega_{ij}(p) \right\rangle
.
\end{split}
\end{equation} 
They are equivalent in the non-interacting case but different in the interacting case.
The field strength $F_1$ is the averaged value and $F_2$ is the quantum fluctuation of the Berry curvature.
The simulation results are shown in Figs.~\ref{figB1} and \ref{figF1}.
The averaged value $F_1$ is zero because of the cancellation between positive and negative contributions.
This is consistent with the non-interacting case.
On the other hand, the fluctuation $F_2$ is nonzero.
The physical value is estimated as $F_2/\tilde{a}^4 \sim 6/\tilde{a}^4 \sim (0.5 \ {\rm GeV})^{-4}$.
The fluctuation seems independent of momentum within the statistical errors.
The momentum independence can be interpreted that gluon scattering gives random phase fluctuation to the fermion wave function.
The randomness is characteristic of the strong gluon interaction, as typified by the chiral random matrix theory.
It is a nontrivial question whether the momentum independence holds in a weakly interacting case.
This would become clear by calculating the fluctuation in perturbation theory.

The above results suggest that, even though the averaged value of the Berry curvature is consistent with the non-interacting value, the strong gluon interaction induces large quantum fluctuation.
The non-interacting value is often used for effective theory of the Berry curvature, such as topological field theory or chiral kinetic theory.
Such a classical picture will be valid only in a weakly coupled regime. 
The quantum fluctuation cannot be negligible in a strongly coupled regime.
Note that this is the simulation at zero temperature.
At a high temperature or density, where the chiral kinetic theory is applied, the gluon interaction is weaker and the quantum fluctuation will be smaller.
Also, in the interacting case, the investigation of the Berry curvature beyond the adiabatic approximation will be needed and may be done in the future.

\section*{Acknowledgments}
S.P.~was supported by a JSPS postdoctoral fellowship for foreign researchers (Grant No.~JP16F16320).
A.Y.~was supported by JSPS KAKENHI (Grant No.~JP15K17624).   
The numerical simulation was carried out on SX-ACE in Osaka University

\appendix

\section{Analytical calculation in the non-interacting case}

Here we would like to derive the Berry connection and curvature for non-interacting continuum fermions.
Similar calculations can be found in Ref.~\cite{Chen:2013iga}. 

We take the gamma matrices in the Dirac basis, 
\begin{equation}
\gamma^{0}=\left(\begin{array}{cc}
1 & 0\\
0 & -1
\end{array}\right),\quad\gamma^{i}=\left(\begin{array}{cc}
0 & \sigma_{i}\\
-\sigma_{i} & 0
\end{array}\right),\;\gamma^{5}=\left(\begin{array}{cc}
0 & 1\\
1 & 0
\end{array}\right),
\end{equation}
where $\sigma_{i}$ is the Pauli matrix. 
The non-interacting Dirac Hamiltonian is 
\begin{equation}
\H
=\left(\begin{array}{cc}
m & \boldsymbol{\sigma}\cdot\mathbf{p}\\
\boldsymbol{\sigma}\cdot\mathbf{p} & -m
\end{array}\right).
\end{equation} 
We will use $p = \sqrt{\mathbf{p}\cdot\mathbf{p}}$ and introduce the polar angles of $\mathbf{p}$,
\begin{equation}
\mathbf{p}=p(\sin\theta\cos\varphi,\sin\theta\sin\varphi,\cos\theta).
\end{equation}
The eigenstates of $\H$ are given by,
\begin{equation}
\Phi_{+} = \sqrt{\frac{E+m}{2E}}\left(
\begin{array}{c}
u_{+}\\
\frac{p}{E+m}u_{+}
\end{array}\right),\;\Phi_{-}=\sqrt{\frac{E+m}{2E}}\left(
\begin{array}{c}
u_{-}\\
-\frac{p}{E+m}u_{-}
\end{array}\right)
\end{equation}
for the positive energy $E=\sqrt{p^{2}+m^{2}}$, and
\begin{equation}
\Psi_{+} = \sqrt{\frac{|E|+m}{2|E|}}\left(
\begin{array}{c}
-\frac{p}{|E|+m}u_{+}\\
u_{+}
\end{array}\right),\;\Psi_{-}=\sqrt{\frac{|E|+m}{2|E|}}\left(
\begin{array}{c}
\frac{p}{|E|+m}u_{-}\\
u_{-}
\end{array}\right)
\end{equation}
for the negative energy $E=-\sqrt{p^{2}+m^{2}}$.
The two-component vectors
\begin{equation}
u_{+}=\left(\begin{array}{c}
e^{-i\varphi}\cos\frac{\theta}{2}\\
\sin\frac{\theta}{2}
\end{array}\right),\;u_{-}=\left(\begin{array}{c}
-e^{-i\varphi}\sin\frac{\theta}{2}\\
\cos\frac{\theta}{2}
\end{array}\right),
\end{equation}
 are the eigenstates of the helicity operator $\boldsymbol{\sigma}\cdot\mathbf{p}$,
\begin{equation}
\boldsymbol{\sigma}\cdot\mathbf{p}u_{\pm}=\pm p u_{\pm}.
\end{equation}

In our discussion, we will concentrate on the particle states $\Phi_{+}$ and $\Phi_{-}$. 
After some calculations, we find
\begin{equation}
\mathbf{A}
=-i\left(\begin{array}{cc}
\Phi^{\dagger}_+\partial_{\mathbf{p}}\Phi_+ & \Phi^{\dagger}_-\partial_{\mathbf{p}}\Phi_+\\
\Phi^{\dagger}_+\partial_{\mathbf{p}}\Phi_- & \Phi^{\dagger}_-\partial_{\mathbf{p}}\Phi_-
\end{array}\right)
=-\frac{1}{2p}\left(\begin{array}{cc}
\mathbf{e}_{\phi}\cot\frac{\theta}{2} & \frac{m}{E}\mathbf{e}_{+}\\
\frac{m}{E}\mathbf{e}_{-} & \mathbf{e}_{\phi}\tan\frac{\theta}{2}
\end{array}\right),\label{eq:Bc_01}
\end{equation}
where $\partial_{\mathbf{p}} = \partial/\partial \mathbf{p}$ and $\mathbf{e}_{\pm}=-(\mathbf{e}_{\phi}\pm i\mathbf{e}_{\theta}).$
It can be decomposed into the U(1) and SU(2) parts,
\begin{equation}
\mathbf{A}=\frac{1}{2}\mathbf{A}^{0} + \sum_{a\ne 0} \frac{1}{2}\sigma^{a} \mathbf{A}^{a}
\end{equation}
with
\begin{align}
\mathbf{A}^{0}=-\frac{1}{p\sin\theta}\mathbf{e}_{\phi},
\quad
\mathbf{A}^{1}=\frac{1}{p}\frac{m}{E}\mathbf{e}_{\phi},
\quad
\mathbf{A}^{2}=-\frac{1}{p}\frac{m}{E}\mathbf{e}_{\theta},
\quad 
\mathbf{A}^{3}=-\frac{1}{p}\cot\theta\mathbf{e}_{\phi}.\label{eq:Bc_02}
\end{align}
Inserting the Berry connection into the Berry curvature \eqref{eqOmega} yields
\begin{equation}
\boldsymbol{\Omega}^{0}= 0,
\quad
\boldsymbol{\Omega}^{1}=\frac{m}{E^{3}}\mathbf{e}_{\theta},
\quad
\boldsymbol{\Omega}^{2}=\frac{m}{E^{3}}\mathbf{e}_{\phi},
\quad
\boldsymbol{\Omega}^{3}=\frac{1}{E^{2}}\mathbf{e}_{p},
\label{eq:Bcur_01}
\end{equation}
where $\Omega_i^\alpha \equiv \epsilon_{ijk}\Omega_{jk}^\alpha$.
By using the Eq.~(\ref{eq:Bcur_01}), we obtain
\begin{align}
\frac{1}{4}\boldsymbol{\Omega}^{0} \cdot \boldsymbol{\Omega}^{0} & =0,\\
\sum_{a\ne 0} \frac{1}{4} \boldsymbol{\Omega}^{a}\cdot\boldsymbol{\Omega}^{a}&=\frac{2m^{2}+E^{2}}{4E^{6}}.
\end{align}

\bibliographystyle{elsarticle-num}
\bibliography{paper}

\end{document}